\documentclass[twocolumn,superscriptaddress,
showpacs,preprintnumbers,amsmath,amssymb]{revtex4}

\usepackage{graphicx}
\usepackage{amssymb,amsmath,amsthm}
\newtheorem{Thm}{Theorem}

\newtheorem{Lem}[Thm]{Lemma}
\newtheorem{Prop}[Thm]{Proposition}
\theoremstyle{definition}

\newcommand{\bra}[1]{{\left\langle #1 \right|}}
\newcommand{\ket}[1]{{\left| #1 \right\rangle}}

\begin{document}

\title{Multipartite bound entanglement and multi-setting Bell inequalities}

\author{Dong Pyo Chi}
\affiliation{
 Department of Mathematical Sciences,
 Seoul National University, Seoul 151-742, Korea
}
\author{Kabgyun Jeong}
\affiliation{
 Nano Systems Institute (NSI-NCRC),
 Seoul National University, Seoul 151-742, Korea
 }
\author{Taewan Kim}
\affiliation{
 Department of Mathematical Sciences,
 Seoul National University, Seoul 151-742, Korea
}
\author{Kyungjin Lee}
\affiliation{
 Department of Mathematical Sciences,
 Seoul National University, Seoul 151-742, Korea
}
\author{Soojoon Lee}
\affiliation{
 Department of Mathematics and Research Institute for Basic Sciences,
 Kyung Hee University, Seoul 130-701, Korea
}
\date{\today}

\begin{abstract}
D\"{u}r~[Phys. Rev. Lett. {\bf 87}, 230402 (2001)] constructed
$N$-qubit bound entangled states which violate a Bell inequality for $N\ge 8$,
and his result was recently improved by showing that
there exists an $N$-qubit bound entangled state violating the Bell inequality
if and only if $N\ge 6$~[Phys. Rev. A {\bf 79}, 032309 (2009)].
On the other hand, it has been also shown that
the states which D\"{u}r considered
violate Bell inequalities different from the inequality for $N\ge 6$.
In this paper, by employing different forms of Bell inequalities,
in particular,
a specific form of Bell inequalities with $M$ settings of the measuring apparatus
for sufficiently large $M$,
we prove that
there exists an $N$-qubit bound entangled state
violating the $M$-setting Bell inequality
if and only if $N\ge 4$.
\end{abstract}

\maketitle

Entanglement is one of the most important properties in quantum mechanics,
and provides us with fruitful applications,
as one can see in quantum communication protocols
such as quantum teleportation and quantum key distribution.

There are two kinds of entangled states.
One is the {\em distillable} entangled (DE) state,
and the other is the {\em bound} entangled (BE) state.
While, from several copies of DE states,
some pure entanglement can be distilled by
local quantum operations and classical communication (LOCC),
one cannot extract any pure entanglement from BE states by LOCC.
Nevertheless, it has been shown that
any BE states are useful
in quantum information processing~\cite{HHH,Masanes1,Ishizaka,HHHO1,HHHO2,HPHH,CCKKL,HA}.
Thus, we need to analyze BE states more carefully.

There is another important property in quantum mechanics, called nonlocality,
which can be seen from violation of some conditions
that are satisfied by any local variable theory.
The conditions are known as {\em Bell inequalities}.
Since Werner~\cite{Werner} discovered DE states which can be described by a local hidden variable model,
there have been a lot of research works about the following question:
Does there exist a BE state violating a Bell inequality?

Since there is no BE state in two-qubit system~\cite{HHH97},
the answer is trivially ``No",
and it was known that if a three-qubit state
violates a specific form of the Bell inequality
then it is distillable~\cite{LJK}.
However, D\"{u}r~\cite{Dur} found that for $N\ge 8$
there exist $N$-qubit BE states
which violate a Bell inequality,
and his result was recently improved by showing that
there exists an $N$-qubit bound entangled state violating the Bell inequality
if and only if $N\ge 6$~\cite{LLK}.
On the other hand, it has been also shown that
the states D\"{u}r considered
violate Bell inequalities different from the inequality for $N\ge 7$ in~\cite{KKCO}
and for $N\ge 6$ in~\cite{SSZ}.

In this paper, by using different forms of Bell inequalities,
in particular,
a specific form of Bell inequalities with $M$ settings of the measuring apparatus,
we show that
if any $N$-qubit state violates the inequality then
there exist at least
\begin{equation}
\left\lfloor 2^{N-1} - \frac{M^N\sin^N(\pi/2M)}{2\cos(\pi/2M)}+1\right\rfloor
\label{eq:No_DBS}
\end{equation}
 distillable bipartite splits.
Therefore, for sufficiently large $M$,
we conclude that
at least one $N$-qubit BE state violates the $M$-setting Bell inequality
if and only if $N\ge 4$.

We first consider a Bell inequality with $M$ settings
on the $N$-qubit system, proposed in~\cite{NLP}.
Let $\mathcal{B'}_N^M$ be the Bell operator
defined as
\begin{equation}
{\mathcal{B'}_N^M}\equiv
\sum_{m_1, \ldots, m_N=0}^{M-1}
c_{m_1,\ldots, m_N}
\vec{\sigma}_{m_1}\otimes\cdots\otimes\vec{\sigma}_{m_N},
\label{eq:B_NM1}
\end{equation}
where $\vec{\sigma}_{m_n}=\sigma_x\cos(\phi_{m_n})+\sigma_y\sin(\phi_{m_n})$,
and the coefficients $c_{m_1, \ldots, m_N}$ are in a form
\begin{equation}
c_{m_1, \ldots, m_N}=\frac{\sin^N(\pi/2M)}{\cos(\pi/2M)}\cos\left(\sum_{j=1}^N\phi_{m_j}\right)
\label{eq:Coeff}
\end{equation}
with the angles given by
\begin{equation}
\phi_{m_n}=\frac{\pi}{M}m_n + \frac{\pi}{2MN}\eta.
\label{eq:phi_mn}
\end{equation}
In Eq.~(\ref{eq:phi_mn}),
the number $\eta=1, 2$ is fixed for a given
experimental situation,
that is,
\begin{equation}
\eta=[M+1]_2[N]_2+1,
\label{eq:ETA}
\end{equation}
where $[x]_2$ stands for $x$ modulo $2$.
Then the $M$-setting Bell inequality is as follows:
\begin{equation}
\left|{\mathrm{tr}\left(\mathcal{B'}_N^M\rho\right)}\right|\leq 1.
\label{eq:B_BD}
\end{equation}

It was shown~\cite{NLP} that
the Bell operator ${\mathcal{B'}_N^M}$ has only two
eigenvalues $\pm ({M^N}/2)\sin^N(\pi/2M)/{\cos(\pi/2M)}$, and
is essentially equivalent to
\begin{equation}
\mathcal{B}_N^M=\frac{M^N\sin^N(\pi/2M)}{2\cos(\pi/2M)}
\left(\ket{\Psi_0^+}\bra{\Psi_0^+}-\ket{\Psi_0^-}\bra{\Psi_0^-}\right),
\label{eq:B_NM2}
\end{equation}
where $\ket{\Psi_0^\pm}$ are $N$-qubit maximally entangled states defined as
\begin{equation}
\ket{\Psi_0^\pm}=\frac{1}{\sqrt{2}}
\left(\ket{00\cdots 0}\pm\ket{11 \cdots 1}\right).
\label{eq:Phi_0}
\end{equation}
Hence, we here deal with the $M$-setting Bell inequality
with respect to the Bell operator in~(\ref{eq:B_NM2}),
$\left|{\mathrm{tr}\left(\mathcal{B}_N^M\rho\right)}\right|\leq 1$.

In order to obtain our results,
we now consider the family of $N$-qubit states $\rho_N$ presented
in~\cite{DCT,DC},
\begin{eqnarray}
\rho_N &=& \sum_{\sigma=\pm}\lambda_0^\sigma\ket{\Psi_0^\sigma}\bra{\Psi_0^\sigma}
\nonumber\\
&&+\sum_{j=1}^{2^{N-1}-1}\lambda_j
\left(\ket{\Psi_j^+}\bra{\Psi_j^+}+\ket{\Psi_j^-}\bra{\Psi_j^-}\right),
\label{eq:rho_N}
\end{eqnarray}
where
\begin{equation}
\ket{\Psi_j^\pm}=\frac{1}{\sqrt{2}}
\left(\ket{j}\ket{0}\pm\ket{2^{N-1}-j-1}\ket{1}\right),
\label{eq:Phi_j}
\end{equation}
and $\lambda_0^+ +\lambda_0^- +2\sum_j \lambda_j =1$.
We remark that any arbitrary $N$-qubit state can be depolarized to a state in this family~\cite{DCT}.
In other words,
by the depolarizing process,
any $N$-qubit state $\rho$ can be transformed into
one in the family of $\rho_N$
with
\begin{eqnarray}
\lambda_0^\pm
&=&\bra{\Psi_0^\pm}\rho\ket{\Psi_0^\pm}=\bra{\Psi_0^\pm}\rho_N\ket{\Psi_0^\pm},
\nonumber\\
2\lambda_j
&=&\bra{\Psi_j^+}\rho\ket{\Psi_j^+}+\bra{\Psi_j^-}\rho\ket{\Psi_j^-}
\nonumber\\
&=& \bra{\Psi_j^+}\rho_N\ket{\Psi_j^+}+\bra{\Psi_j^-}\rho_N\ket{\Psi_j^-}.
\label{eq:depolarization}
\end{eqnarray}



For each  $0<j<2^{N-1}$,
let $P_j$ be the bipartite split such that
the coefficient of $2^{N-i-1}$ in the binary representation of $j$ is zero
if and only if party $i$ belongs to the same set as the last party.
Then the following proposition about bipartite distillability of the states $\rho_N$
has been known by D\"{u}r and Cirac~\cite{DC}.
\begin{Prop}\label{Prop:DC}
$\rho_N$ is distillable for the bipartite split $P_j$
if and only if  $2\lambda_j<\Delta\equiv\lambda_0^+ -\lambda_0^-$.
\end{Prop}

By exploiting the proof of Lemma~2 in Ref.~\cite{LLK} and Proposition~\ref{Prop:DC},
we can obtain the following key lemma for our results.
\begin{Lem}\label{Lem:main}
If
\begin{equation}
\Delta > \frac{2\cos(\pi/2M)}{M^N\sin^N(\pi/2M)}\label{eq:violation}
\end{equation}
then there exist
at least $\lfloor 2^{N-1} - \frac{M^N\sin^N(\pi/2M)}{2\cos(\pi/2M)}+1\rfloor$
distillable bipartite splits in $\rho_N$.
\end{Lem}
\begin{proof}
Let $m$ be the number of distillable bipartite splits,
$P_{j_1}, P_{j_2}, \ldots , P_{j_m}$.
Suppose that $m \le 2^{N-1} - \frac{M^N\sin^N(\pi/2M)}{2\cos(\pi/2M)}$.
Then we readily obtain the following inequality: 
\begin{eqnarray}
1-\Delta 
&\ge& 
2\sum_{j=1}^{2^{N-1}-1} \lambda_j \nonumber \\
&=& 2(\lambda_{j_1}+\lambda_{j_2}+\cdots+ \lambda_{j_m})
+2\sum_{j\notin \{j_1,\ldots,j_m\}} \lambda_j \nonumber \\
&\ge& 2(\lambda_{j_1}+\lambda_{j_2}+\cdots+ \lambda_{j_m}) 
+(2^{N-1}-1-m)\Delta.\nonumber \\
\label{eq:ineqs01}
\end{eqnarray}
It follows that
\begin{eqnarray}
1&\ge&  2(\lambda_{j_1}+\lambda_{j_2}+\cdots+ \lambda_{j_m})
+(2^{N-1}-m)\Delta\nonumber \\
&>&  2(\lambda_{j_1}+\lambda_{j_2}+\cdots+ \lambda_{j_m})
+\frac{(2^{N-1}-m)}{\left(\frac{M^N\sin^N(\pi/2M)}{2\cos(\pi/2M)}\right)} \nonumber \\
&\ge& 2(\lambda_{j_1}+\lambda_{j_2}+\cdots+ \lambda_{j_m}) +1.
\label{eq:ineqs02}
\end{eqnarray}
The inequality~(\ref{eq:ineqs02})
leads to a contradiction. 
Therefore, we can conclude that $m> 2^{N-1} - \frac{M^N\sin^N(\pi/2M)}{2\cos(\pi/2M)}$.
\end{proof}

Then, by equalities in~(\ref{eq:depolarization}),
we obtain the following equalities:
\begin{eqnarray}
\frac{2\cos({\pi}/{2M})}{M^N\sin^N({\pi}/{2M})}\mathrm{tr}\left(\mathcal{B}_{N}^M\rho\right)
&=&
\bra{\Psi_0^+}\rho\ket{\Psi_0^+}-\bra{\Psi_0^-}\rho\ket{\Psi_0^-}
\nonumber\\
&=&
\bra{\Psi_0^+}\rho_{N}\ket{\Psi_0^+}-\bra{\Psi_0^-}\rho_{N}\ket{\Psi_0^-}
\nonumber\\
&=&
\lambda_0^+ - \lambda_0^- =\Delta,
\label{eq:lambda_Delta}
\end{eqnarray}
where $\rho$ is a given arbitrary $N$-qubit state,
and $\rho_N$ is the state transformed from $\rho$ by the depolarizing process.
Hence, we have the following theorem by Lemma~\ref{Lem:main}.
\begin{Thm}\label{Thm:main}
For all the $N$-qubit states $\rho$
violating the $M$-setting Bell inequality
with respect to the Bell operator in~(\ref{eq:B_NM2}),
there exist
at least $\lfloor 2^{N-1} - \frac{M^N\sin^N(\pi/2M)}{2\cos(\pi/2M)}+1\rfloor$ distillable bipartite splits.
\end{Thm}

Theorem~\ref{Thm:main} provides us with
a necessary and sufficient condition for
the existence of $N$-qubit BE states violating the $M$-setting Bell inequality
with respect to the Bell operator in~(\ref{eq:B_NM2})
for sufficiently large $M$.
In order to show the condition,
we begin with reminding the following proposition
about a relation between distillability and negative partial transposition (NPT),
which has been shown by D\"{u}r and Cirac~\cite{DCT}.
\begin{Prop}\label{Prop:DCT}
A maximally entangled pair between particles $i$ and $j$
can be distilled from $\rho_N$ if and only if
all possible bipartite splits of $\rho_N$
where the particles $i$ and $j$ belong to different parties,
have NPT.
\end{Prop}

By Theorem~\ref{Thm:main} and Proposition~\ref{Prop:DCT},
we can prove the following theorem.
\begin{Thm}\label{Thm:main2}
For sufficiently large $M$,
there exists at least one $N$-qubit BE state
violating the $M$-setting Bell inequality
with respect to the Bell operator in~(\ref{eq:B_NM2})
if and only if $N\ge 4$.
\end{Thm}
\begin{proof}
We note that the number of total bipartite splits is $2^{N-1}-1$,
and that the number of all distillable bipartite splits is
at least $\lfloor 2^{N-1} - \frac{M^N\sin^N(\pi/2M)}{2\cos(\pi/2M)}
+1\rfloor$ by Theorem~\ref{Thm:main}.

We first assume that $N=3$.
Then it follows from Theorem~\ref{Thm:main} that all bipartite splits are distillable,
and so have NPT.
By Proposition~\ref{Prop:DCT},
a maximally entangled state can be distilled
between any particles $i$ and $j$.

Conversely, if $N\ge 4$ then
the $N$-qubit state $\varrho_N$ presented in Ref.~\cite{LLK}
violates the $M$-setting Bell inequality
for sufficiently large $M$, as follows:
The $N$-qubit state $\varrho_N$ is defined as
\begin{eqnarray}
\varrho_N&=&\frac{1}{N-1}\ket{\Psi_0^+}\bra{\Psi_0^+}
\nonumber \\
&&+\frac{1}{2(N-1)}\sum_{j\in J_N}
\left(\ket{\Psi_j^+}\bra{\Psi_j^+}+\ket{\Psi_j^-}\bra{\Psi_j^-}\right),
\nonumber \\
\label{eq:BErho_N}
\end{eqnarray}
where $J_N=\{3, 6, \ldots, 3\cdot 2^{N-3}\}$.
Then, for $M\ge 6$,
we can readily obtain that
\begin{equation}
\left|{\mathrm{tr}\left(\mathcal{B}_N^M\varrho_N\right)}\right|
=\frac{M^N\sin^N(\pi/2M)}{2(N-1)\cos(\pi/2M)}>1
\label{eq:BMN_rho_N}
\end{equation}
if and only if $N\ge 4$,
as seen in Fig.~\ref{Fig:BMN}.
\begin{figure}
\includegraphics[angle=-90,width=.95\linewidth]{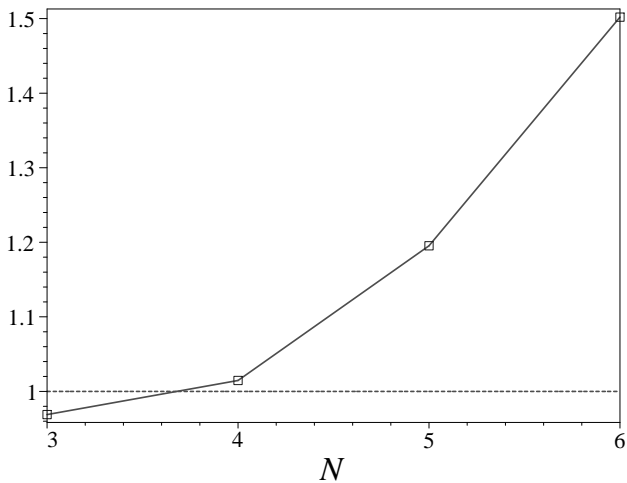}
\caption{\label{Fig:BMN}
The expectation value of the $M$-setting Bell operator $\mathcal{B}_N^M$ in~(\ref{eq:B_NM2})
for the state $\varrho_N$, that is,
$\left|{\mathrm{tr}\left(\mathcal{B}_N^M\varrho_N\right)}\right|$: $M$ is sufficiently large ($M\ge 6$).}
\end{figure}
Therefore, the state $\varrho_N$ violates the $M$-setting Bell inequality
with respect to the Bell operator in~(\ref{eq:B_NM2}).

Furthermore,
by the same reason as that in Ref.~\cite{LLK},
it can be shown that
the $N$-qubit state $\varrho_N$ is undistillable,
and hence there exists an $N$-qubit BE state $\varrho_N$
violating the $M$-setting Bell inequality if $N\ge 4$ for sufficiently large $M$.
%
\end{proof}
Remark that $\left|{\mathrm{tr}\left(\mathcal{B}_N^M\varrho_N\right)}\right|$
in the inequality~(\ref{eq:BMN_rho_N}) increases as $M$ tends to infinity,
but if $M \geq 6$ then we have the same result as in Theorem~\ref{Thm:main2}
for any $M$-setting Bell inequality,
since
\begin{equation}
\lim_{M \rightarrow \infty}\left|{\mathrm{tr}\left(\mathcal{B}_N^M\varrho_N\right)}\right|
= \frac{\pi^N}{2^{N+1}(N-1)}>1
\label{eq:BMN_infty}
\end{equation}
if and only if $N \geq 4$.

In conclusion,
by employing a specific form of Bell inequalities with $M$ settings of the measuring apparatus
for sufficiently large $M$,
we have shown that
if any $N$-qubit state violates the inequality then
there exist at least
$\left\lfloor 2^{N-1} - \frac{M^N\sin^N(\pi/2M)}{2\cos(\pi/2M)}+1\right\rfloor$
distillable bipartite splits,
and have concluded that
there exists an $N$-qubit BE state
violating the $M$-setting Bell inequality
if and only if $N\ge 4$.

This work improves the previous results~\cite{Dur,KKCO,SSZ,LLK}
related to the multipartite BE states and Bell inequalities,
even though it has been already known that
there exists a four-qubit BE state, the so-called Smolin state~\cite{Smolin},
violating some other Bell inequality~\cite{AH}.

\begin{figure}
\includegraphics[width=.8\linewidth]{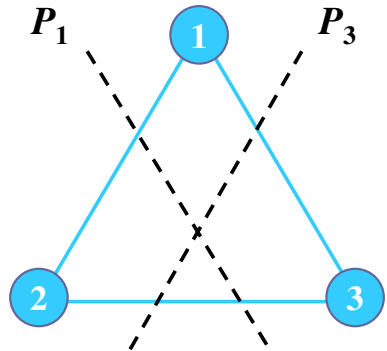}
\caption{\label{Fig:rho3}
A three-qubit BE state $\varrho'_3$ violating the positive partial transpose inequality in~\cite{Nagata}:
$P_1$ and $P_3$ are undistillable bipartite splits.
}
\end{figure}

Furthermore, our technique in this paper can be also applied to
the positive partial transpose inequality,
$\left|{\mathrm{tr}\left(\mathcal{P}_N\rho\right)}\right| \le 1$ with
$\mathcal{P}_N=2^{N-1}\left(\ket{\Psi_0^+}\bra{\Psi_0^+}-\ket{\Psi_0^-}\bra{\Psi_0^-}\right)$,
which was proposed in~\cite{Nagata},
and so we can construct a three-qubit BE state $\varrho'_3$
violating the inequality as in Fig.~\ref{Fig:rho3}:
\begin{eqnarray}
\varrho'_3=\frac{1}{3}\ket{\Psi_0^+}\bra{\Psi_0^+}
+\frac{1}{6}\sum_{j\in \{1, 3\}}
\left(\ket{\Psi_j^+}\bra{\Psi_j^+}+\ket{\Psi_j^-}\bra{\Psi_j^-}\right),
\nonumber \\
\label{eq:BErho_3}
\end{eqnarray}
since $\left|{\mathrm{tr}\left(\mathcal{P}_3\varrho'_3\right)}\right|=4/3>1$.
Hence, the result in~\cite{Nagata} can be enhanced as well.

This work was supported by the IT R\&D program of MKE/IITA
(2008-F-035-02,
Development of key technologies for commercial quantum cryptography communication system).
D.P.C. was supported by the Korea Science and Engineering Foundation
(KOSEF) grant funded by the Korea government (MOST) (No.~R01-2006-000-10698-0),
and S.L. was supported by the Korea Research Foundation Grant funded by the Korean Government
(MOEHRD, Basic Research Promotion Fund) (KRF-2007-331-C00049).


\end{document}